\documentclass[aps,prb,twocolumn,superscriptaddress,showpacs,10pt]{revtex4-1}
\usepackage{graphicx,amssymb,bm,natbib,amsfonts,amsmath}
\usepackage{float}

\begin{document}

\title{High harmonic generation in undoped graphene: Interplay of inter- and intraband dynamics}
\author{Ibraheem Al-Naib}
\affiliation{Department of Physics, Engineering Physics and Astronomy,
Queen's University, Kingston, Ontario K7L 3N6, Canada}
\author{J. E. Sipe}
\affiliation{Department of Physics and Institute for Optical Sciences,
University of Toronto, Toronto, Ontario M5S 1A7, Canada }
\author{Marc M. Dignam}
\affiliation{Department of Physics, Engineering Physics and Astronomy,
Queen's University, Kingston, Ontario K7L 3N6, Canada}

\begin{abstract}
% 600 characters with spaces!!
We develop a density matrix formalism in the length gauge to calculate the
nonlinear response of intrinsic monolayer graphene at terahertz frequencies.
Employing a tight-binding model, we find that the interplay of the interband
and intraband dynamics leads to strong harmonic generation at moderate field
amplitudes. In particular, we find that at low temperature, the reflected
field of undoped suspended graphene exhibits a third harmonic amplitude that
is 32\% of the fundamental for an incident field of 100 V/cm. Moreover, we find that up to the seventh harmonic and beyond are generated.
\end{abstract}

\maketitle
\section{Introduction}
As the paradigmatic example of a system exhibiting zero-gap Dirac points and
linear band dispersion \cite{Wallace1947,DasSarma2011}, graphene has been
the subject of a host of studies, many with a focus on its fundamental
physical and chemical properties, and many with a view towards device
applications \cite{Novoselov2004,
Geim2009,Schwierz2010,Strait2011,Levesque2011,Weis2012,Mak2012,
Sensale-Rodriguez2012a,Kumar2013,Hong2013,Tassin2013,Glazov2014}. The
absorption of electromagnetic radiation in graphene is controlled by
interband and intraband transitions \cite{DasSarma2011}, as schematically
shown in Fig. 1(a). In undoped graphene, applied fields lead to interband
dynamics by inducing transitions between the bands, while the subsequent
driving of the carriers within their bands by those same fields leads to
intraband dynamics. Transitions near the Dirac point are accessed by
terahertz (THz) fields, while for the intraband dynamics to result in
significant currents the fields must have a relatively low frequency, as do
THz waves \cite{Tonouchi2007,Jepsen2011,Docherty2012}.

Thus while the recent development of sources of intense THz pulses has led
to new avenues of exploration across all of condensed matter physics \cite%
{Hebling2002,Tanaka2011,Hoffmann2011,Al-Naib2013}, terahertz studies should
be singularly suitable for revealing the physics of carrier dynamics in
graphene. A typical THz single cycle pulse and its spectrum are shown in
Figs. 1(b) and 1(c), respectively. Earlier theoretical studies anticipated
strong optical nonlinearity in graphene, leading to effects such as high
harmonic generation \cite%
{Mikhailov2007,Mikhailov2008,Wright2009,Ishikawa2010}. These estimates
relied on the extreme nonparabolicity of the linear dispersion relation to
lead to strong nonlinearity in the intraband response. Yet although a
recent experimental study has demonstrated generation of a third harmonic on
the order of $10^{-3}$ of the transmitted field power for a 45-layer
graphene sample \cite{Bowlan2014}, there has been no experimental indication
of such nonlinearities at terahertz frequencies using \emph{monolayer}
graphene \cite{Paul2013}.
\begin{figure}[b]
\includegraphics[width=8.5cm]{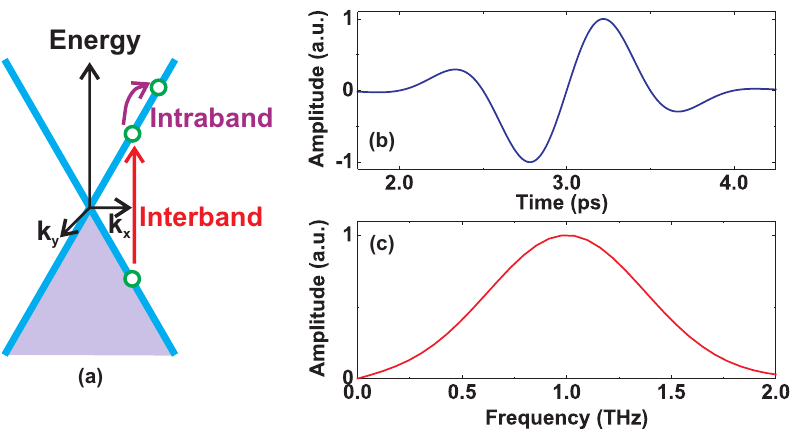}
\caption{(color online) (a) Schematic band structure of undoped $(\protect%
\mu _{F}=0)$ graphene near the Dirac point demonstrating interband and the
intraband dynamics, (b) temporal plot of the incident pulse employed in the
simulations, and (c) the corresponding amplitude spectrum of the pulse.}
\end{figure}
In this paper, we calculate that graphene can indeed exhibit a strong
nonlinear response to THz radiation, with a strong third harmonic field
emitted at incident fields as low as 100 V/cm, along with emission up to the 
$7^{th}$ harmonic and beyond. However, the strong nonlinear response is
not due primarily to the intraband motion of the carriers, but rather to an
interplay between the interband and intraband motion of the carriers. The
critical requirements for such strong nonlinear effects are that the sample
is undoped, i.e., the chemical potential is at the Dirac point
and that the experiments are performed well below room temperature at 10 K.
This low density, low temperature condition has not been investigated
experimentally to date and represents a different regime for harmonic generation
in graphene.
\section{Theory}
Our calculations are based on a theoretical approach employing a
density-matrix formalism in the length gauge. We use a
nearest-neighbor tight-binding model to treat the $\pi-$electrons in the
graphene \cite{Reich2002,DasSarma2011}. The tight binding expression for the Bloch states is given by: 
\begin{equation}
\phi _{n\mathbf{k}}(\mathbf{r})=N\sum_{\mathbf{R}}[\varphi _{pz}(\delta 
\mathbf{r}_{A})+C\varphi _{pz}(\delta \mathbf{r}_{B})]e^{i\mathbf{k}\cdot 
\mathbf{R}},
\end{equation}%
{\setlength{\parindent}{0cm}where $N$ is a normalization factor, }$n=\left\{
c,v\right\}$ labels the conduction and valence bands and {the sum is over
the Bravais lattice vectors $\mathbf{R}$. The factor, $C=\sigma
_{n}e^{-i\chi \left( \mathbf{k}\right) }$, where $\sigma _{v}=1$ and $\sigma
_{c}=-1$ and $\chi \left( \mathbf{k}\right) =\arg \left[ F(\mathbf{k})\right]
$ is the phase between the two sublattice states, where $F\left( \mathbf{k}%
\right) \equiv 1+e^{-i\mathbf{k}\cdot \mathbf{a}_{1}}+e^{-i\mathbf{k}\cdot 
\mathbf{a}_{2}}$, with $\mathbf{a}_{1}$ and $\mathbf{a}_{2}$ being the
primitive Bravais lattice translation vectors. The $\varphi _{pz}(\mathbf{r}%
)$ are the $2p_{z}$ orbitals of carbon, and $\mathbf{\delta }\mathbf{r}%
_{A,B}\equiv \mathbf{r}-\mathbf{R}-\mathbf{r}_{A,B}$, where $\mathbf{r}_{A},%
\mathbf{r}_{B}$ are the basis vectors. In what follows, to simplify the
notation we take the origin in }$k$-space to be at {$\mathbf{K}$-Dirac point
(where $F\left( \mathbf{k}\right) =0$). Near the Dirac point, it is easy to
show that $\chi \left( \mathbf{k}\right) \simeq \theta +\pi /2$ where $%
\theta$ is the angle that $\mathbf{k}$ makes with the $k_{x}$-axis. In
addition, the dispersion for the two bands is given approximately by $%
E_{n}\left( \mathbf{k}\right) \simeq E_{pz}-\sigma _{n}\mathrm{v}_{F}\hslash
k$, where $\mathrm{v}_{F}$ is Fermi velocity {and $E_{pz}$ is the energy of
the $2p_{z}$ states in carbon.} }

It has been shown that the nonlinear response of semiconductors can only be
reliably treated in a two band model if one employs the length gauge \cite%
{Aversa1995,Virk2007}. In particular, if the more common velocity gauge is
used, unphysical divergences arise in the nonlinear response at low
frequencies, such as in the THz range, that can only be removed by
developing sum rules; these become extremely complicated if one works to
high order in the field \cite{Virk2007}. {{Thus, in contrast to previous theoretical work on the nonlinear THz response of graphene~\cite{Mikhailov2007,Mikhailov2008,Wright2009,Ishikawa2010}, we employ here the length-gauge Hamiltonian, given by $H=H_{0}-e\mathbf{r}\cdot \mathbf{E}(t)$, 
%\begin{equation}
%H=H_{0}-e\mathbf{r}\cdot \mathbf{E}(t),
%\end{equation}%
}where $H_{0}$ is the full Hamiltonian of unperturbed graphene, $e=-|e|$ is
the charge of an electron, $\mathbf{r}$ is the electron position vector and $%
\mathbf{E}(t)$ is the THz electric field at the graphene. The carrier
dynamics are calculated by solving the equations of motion for the reduced
density matrix in the basis of conduction band and valence band Bloch
states. These equations require the matrix elements of the
Hamiltonian, which are given by 
\begin{equation}
\text{$\left\langle n\mathbf{k}\right\vert $}H\left\vert \text{$m\mathbf{
k^{\prime }}$}\right\rangle =E_{n}\left( \mathbf{k}\right) \delta
_{nm}\delta \left( \mathbf{k-k}^{\prime }\right) -e\text{$\left\langle n
\mathbf{k}\right\vert $}\mathbf{r}\left\vert \text{$m\mathbf{k^{\prime }}$}
\right\rangle \cdot \mathbf{E}(t).  \label{EqHmat}
\end{equation}%
}

The matrix elements of $\mathbf{r}$ between Bloch states can be shown to be
given by \cite{Aversa1995,Blount1962}: 
\begin{equation}
\text{$\left\langle n\mathbf{k}\right\vert $}\mathbf{r}\left\vert \text{$m%
\mathbf{k^{\prime }}$}\right\rangle =\delta (\mathbf{k}-\mathbf{k^{\prime }})%
\mathbf{\xi }_{nm}\left( \mathbf{k}\right) +i\delta _{nm}\mathbf{\nabla }_{%
\mathbf{k}}\delta (\mathbf{k}-\mathbf{k^{\prime }}), \label{eq_3}
\end{equation}%
{where the connection elements, }$\mathbf{\xi }_{nm}\left( \mathbf{k}\right) 
$,{\ are given by} 
\begin{equation}
\mathbf{\xi }_{nm}\left( \mathbf{k}\right) =i\frac{1}{\Omega _{c}}\int d^{3}%
\mathbf{r}u_{n,\mathbf{k}}^{\ast }\left( \mathbf{r}\right) \mathbf{\nabla }_{%
\mathbf{k}}u_{m,\mathbf{k}}\left( \mathbf{r}\right) ,
\end{equation}%
where $\Omega _{c}$ is the area of a unit cell and $u_{n\mathbf{k}}\left( 
\mathbf{r}\right) $ is the periodic part of the Bloch function. We have
evaluated these connection elements using our tight-binding wavefunction.
Ignoring the overlap of atomic wavefunctions on different atoms, near the
Dirac point the connection elements are given approximately by 
\begin{equation}
\mathbf{\xi }_{nm}\left( \mathbf{k}\right) =\left[ 2\delta _{nm}-1\right] 
\frac{\mathbf{\widehat{\theta }}}{2k}.  \label{zeta}
\end{equation}

To deal with the derivative appearing in Eq.~\ref{eq_3}, as it enters our equations below, we discretize in $k$-space; that
discretization should be understood for the other derivatives of quantities in $k$-space that appear below. Due to the large energy barriers between the $\mathbf{K}$ and $\mathbf{K^{\prime}}$ Dirac cones and the small energies of the THz photons, we can treat the dynamics of the electrons near
the two Dirac points as being disconnected and simply sum their (identical)
contributions to the current density. We define the reduced density matrix
elements to be 
%\begin{equation}
$\rho _{nm}\left( \mathbf{k}\right) \equiv \left\langle a_{m,\mathbf{k}}^{\dag }a_{n,\mathbf{k}}\right\rangle$, %  \label{Eqrhonm}
%\end{equation}%
{where} $a_{n,\mathbf{k}}^{\dag }$ $\left( a_{n,\mathbf{k}}\right) $ is the
creation (annihilation) operator for an electron in the Bloch state $%
\left\vert n\text{$\mathbf{k}$}\right\rangle $. Using Eq.~(\ref{EqHmat}) for
the matrix elements of the Hamiltonian, the dynamic equations for the
reduced density matrix elements become: 
\begin{align}
& \frac{d\rho _{cv}\left( \mathbf{k}\right) }{dt}=\frac{ie\mathbf{E}\left(
t\right) \cdot \mathbf{\xi }_{cv}\left( \mathbf{k}\right) }{\hslash }\left[
\rho _{vv}\left( \mathbf{k}\right) -\rho _{cc}\left( \mathbf{k}\right) %
\right]   \notag \\
& \!-i\omega _{cv}\left( \mathbf{k}\right) \rho _{cv}\left( \mathbf{k}%
\right) -\frac{e\mathbf{E}\left( t\right) \cdot \mathbf{\nabla }_{\mathbf{k}%
}\rho _{cv}\left( \mathbf{k}\right) }{\hslash }-\frac{\rho _{cv}\left( 
\mathbf{k}\right) }{\tau },  \label{Eqrhocv}
\end{align}%
%and%
\begin{align}
& \frac{d\rho _{nn}\left( \mathbf{k}\right) }{dt}=-\frac{i\sigma _{n}e\cdot 
\mathbf{E}\left( t\right) }{\hslash }\left[ \mathbf{\xi }_{cv}\left( \mathbf{%
k}\right) \rho _{vc}\left( \mathbf{k}\right) -\mathbf{\xi }_{vc}\left( 
\mathbf{k}\right) \rho _{cv}\left( \mathbf{k}\right) \right]   \notag \\
& \;\;\;\;\;\;-\frac{e\mathbf{E}\left( t\right) \cdot \mathbf{\nabla }_{%
\mathbf{k}}\rho _{nn}\left( \mathbf{k}\right) }{\hslash }-\frac{\left[ \rho
_{nn}\left( \mathbf{k}\right) -f_{n}\left( \mathbf{k},t\right) \right] }{%
\tau _{n}},  \label{Eqrhonn}
\end{align}%
{where $\omega _{cv}\left( \mathbf{k}\right) \simeq 2v_{F}k$ and }$%
f_{n}\left( \mathbf{k},t\right) $ {is a Fermi-Dirac distribution with a
time-dependent temperature}{. Note that the electric field in both equations
is the field at the graphene, which is equal to the amplitude of the
transmitted field.\ In our numerical implementation, we model the vacancy
populations rather than the valence band electrons to allow us to only
include states near the Dirac point. Because the scattering times in graphene are only on
the order of tens of femtoseconds \cite%
{Tse2008,Breusing2009,Lui2010,Tani2012,Tielrooij2013,Paul2013,Bowlan2014}, the inclusion of
scattering processes is an essential element in any model of the THz
response of graphene. For the low carrier densities considered in this work, carrier-carrier scattering is expected to be relatively unimportant, and the dominant scattering processes will be defect scattering and electron-phonon scattering. In the above equations, we treat the scattering
phenomenologically. For the interband coherences, $\rho _{cv}\left( \mathbf{k%
}\right) $, we introduce an interband decoherence time, $\tau $, which we
assume is independent of $\mathbf{k}$. The populations relax back to
Fermi-Dirac thermal distributions, }$f_{n}\left( \mathbf{k},t\right) $,\
with relaxation times, $\tau _{n}$ and a temperature that is chosen to
obtain the carrier populations, which are time-dependent due to the
THz-induced interband transitions. As it has been found both experimentally and theoretically
\cite{Winnerl2011} that the time taken for conduction band electrons to
relax to the valence band is much longer than intraband scattering times, we
neglect interband relaxation.{\ }

To solve the above equations, we employ a direct approach, where we put $%
\mathbf{k}$ on a grid and step through time using a Runge-Kutta algorithm.
In order to facilitate this, we employ balanced difference approximations to
the gradients. Given the geometry of the lattice and Brillouin zone, we
employ a hexagonal grid with a uniform point density in $k$-space. Following
the formalism of Aversa and Sipe \cite{Aversa1995}, the current density is
given by 
\begin{equation}
\mathbf{J}\left( t\right) =\frac{e}{mA}Tr\left\{ \mathbf{p}\widehat{\rho }%
\left( t\right) \right\} =\frac{e}{i\hslash A}Tr\left\{ \left[ \mathbf{r,}H%
\right] \widehat{\rho }\left( t\right) \right\} ,
\end{equation}%
{where the trace is over single-electron states, $A$ is the normalization
area of the graphene sheet, $\mathbf{p}$ is the electron momentum operator
and }$\widehat{\rho }\left( t\right) $ is the reduced density matrix with
matrix elements $\rho _{nm}\left( \mathbf{k}\right)${. Using our expression for the
Hamiltonian and the matrix elements of the position operator, it is possible
to write the current density as the sum of an interband term, $%
\mathbf{J}_{e}$, and an intraband term, $\mathbf{J}_{i}$. After
considerable work, one can show that} {the interband current density is
given by%\cite{factor_of_4} } 
\begin{equation}
\mathbf{J}_{e}\left( t\right) =\frac{8\left\vert e\right\vert }{A}{Re}%
\left\{ \sum_{\mathbf{k}}\frac{\mathbf{\widehat{\theta }}}{2k}\frac{d\rho
_{cv}\left( \mathbf{k},t\right) }{dt}\right\} ,
\end{equation}%
while the intraband current density is given by 
\begin{align}
\mathbf{J}_{i}\left( t\right) & =\frac{-4|e|v_{F}}{A}\sum_{\mathbf{k}%
}\left\{ \rho _{cc}\left( \mathbf{k},t\right) -\rho _{vv}\left( \mathbf{k}%
,t\right) \right\} \widehat{\mathbf{k}}  \notag \\
& \!\!\!+\frac{8|e|^{2}}{A\hslash }\sum_{\mathbf{k}}{Re}\left\{ \rho _{cv}\left( 
\mathbf{k},t\right) \mathbf{\nabla }_{\mathbf{k}}\left[ \mathbf{E}\left(
t\right) \cdot \mathbf{\xi }_{vc}\left( \mathbf{k}\right) \right] \right\}. \label{currents}
\end{align}%
The sums over $\mathbf{k}$ in the current density expressions are restricted to a region near the $\mathbf{K}$-Dirac point and we include a factor of 4 to account both for spin degeneracy and the contributions from carriers from the $\mathbf{K^{\prime}}$-Dirac point. Note that because to first order $\rho _{cv}(\mathbf{k})$ is symmetric under
inversion, the terms on $\mathbf{J}_{i}\left( t\right) $ given on the second
line of Eq.~(\ref{currents}) are to third order and higher in the electric field and are
very small for the conditions considered. We consider a suspended graphene
sample such as employed experimentally by Paul \textit{et al.} \cite%
{Paul2013} and use the time-dependent current sheet densities to calculate
the transmitted and the reflected THz fields. Although we are interested in the nonlinear response, we have verified that we obtain the expected linear conductivities numerically as a check. Furthermore, we have verified convergence in the nonlinear regime by changing the grid density, the extent of the grid, the time-step tolerance and the polarization of the incident field.
\section{Results}
We now present the results of our simulations of the nonlinear THz response
for undoped $(\mu _{F}=0)$ monolayer graphene. We run the simulation for a
temperature of 10 K (corresponding to intrinsic density of $9.32\times
10^{7}/cm^{2}$) so as to minimize initial Pauli-blocking and thus maximize
the interband transitions. The input THz pulse is a sinusoidal Gaussian
pulse with central frequency $\nu$ of 1 THz and duration (FWHM) of 1 ps as shown
in Fig. 1(b). The relaxation times $\tau $ and $\tau _{n}$ have been chosen
to be 50 fs, which is an average of various theoretical and measured values
for such (sample-dependent) constants \cite%
{Tse2008,Breusing2009,Lui2010,Tani2012,Tielrooij2013,Paul2013,Bowlan2014}.
This system has a thermal distribution of carriers with an average electron
energy of 0.86 meV, which is much less than average photon energy of 4.14
meV of the 1 THz pulse. In the simulation, we keep the chemical potential at
the Dirac point but as carriers are injected we calculate the new effective temperature in the
distributions, $f_{n}\left( \mathbf{k},t\right)$, to which the carriers
relax to account for the increase in the carrier density. All of the
carriers (injected as well as thermal) are driven by the applied electric
field (intraband dynamics). This results in the opening and closing of
different states for extra injection of carriers. This interplay between
intraband and interband dynamics is at the heart of the highly nonlinear
response that we now examine. 
\begin{figure}[t!]
\includegraphics[width=8.5cm]{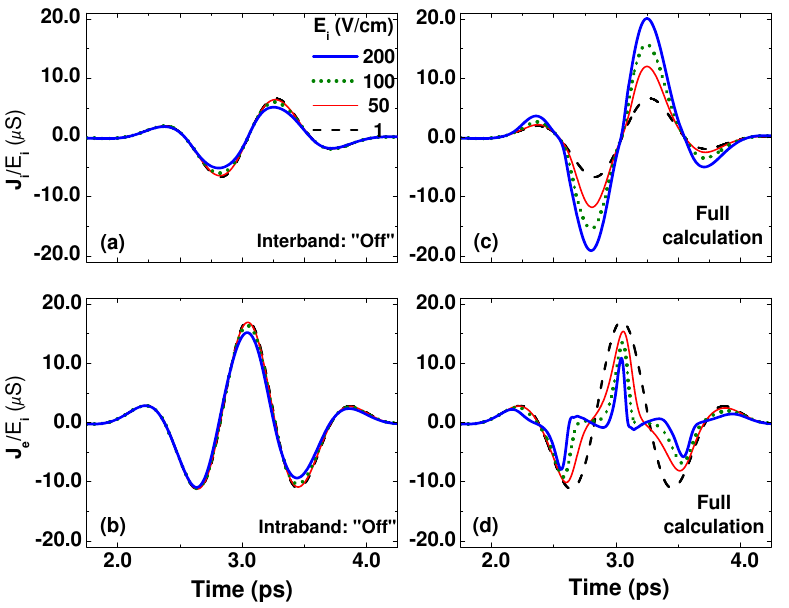}
\caption{(color online) (a) Intraband current in the first case when the ${%
\protect\xi }$'s are set to zero, (b) interband current in the second case
when all the gradients in Eqs. (\ref{Eqrhocv}) and (\ref{Eqrhonn}) are eliminated, (c) intraband and (d)
interband current for the full calculation. In all plots the current is
calculated for the four different incident electric field with peak $E_{i}$
of 1, 50, 100, and 200 V/cm. All currents have been normalized to the
corresponding $E_{i}$.}
\end{figure}

In Fig. 2 we plot the intraband and interband current densities for four different incident field peak amplitudes (1, 50, 100 and 200 V/cm). In order to
facilitate the comparison, all current densities are normalized to the peak, 
$E_{i}$, of the incident field, such that, if the response were linear,
these relative currents would be unchanged by an increase in incident field.
To demonstrate the importance of the interplay of interband and intraband
dynamics, we first present the results of the calculation when the interplay
is missing. In the first case (Fig. 2(a)), we have set the connection
elements ($\xi $'s) to zero in Eqs. (\ref{Eqrhocv}) and (\ref{Eqrhonn}) such
that there is no electron motion between the bands. In this case, only the
intraband current contributes. The relative current is decreased near the
peaks (clipping) when the electric field is increased. This clipping results
in a 23\% reduction in the peak relative current when the field is 200 V/cm
compared to the linear response; the reduction is expected due to the linear
dispersion of graphene \cite{Wallace1947,DasSarma2011}, but the nonlinearity
is quite modest. In the second case (Fig. 2(b)), all of the gradients in
Eqs. (\ref{Eqrhocv}) and (\ref{Eqrhonn}) have been excluded. Interband
transitions occur but the transfer of electrons between bands is not
accompanied by the subsequent motion of electrons within individual bands.
Here the relative interband current is seen to decrease as the THz electric
field increases.\ This is due to interband absorption
saturation that arises due to Pauli-blocking. Again, the nonlinearity is
modest.\ 

Finally, we present the results of the\emph{\ full calculation} that
includes the full interband and intraband carrier dynamics. The relative
intraband and interband current densities for this full calculation are
shown in Figs. 2(c) and 2(d), respectively. The intraband current undergoes
a large increase as the field is increased and is almost tripled when we go
from a field of 1 V/cm to the highest field of 200 V/cm. This increase is
expected, as it arises from the increase in the carrier densities due to the
interband injection of carriers. In addition, there is a relatively small
distortion in the intraband current as the field is increased. Finally, in
Fig. 2(d), we plot the relative interband current density.\ We see first
that as the incident field increases, there is a decrease in the relative
interband current. Most importantly, the temporal form of the interband
current is greatly modified, especially at fields at and above 100 V/cm.\ It
is evident by comparison to Fig. 2(b) that this strong nonlinearity only
arises when both interband an intraband processes are included in the
calculation, and is thus due to the interplay between the interband and
intraband dynamics.%\newline
\begin{figure}[t]
\includegraphics[width=8.cm]{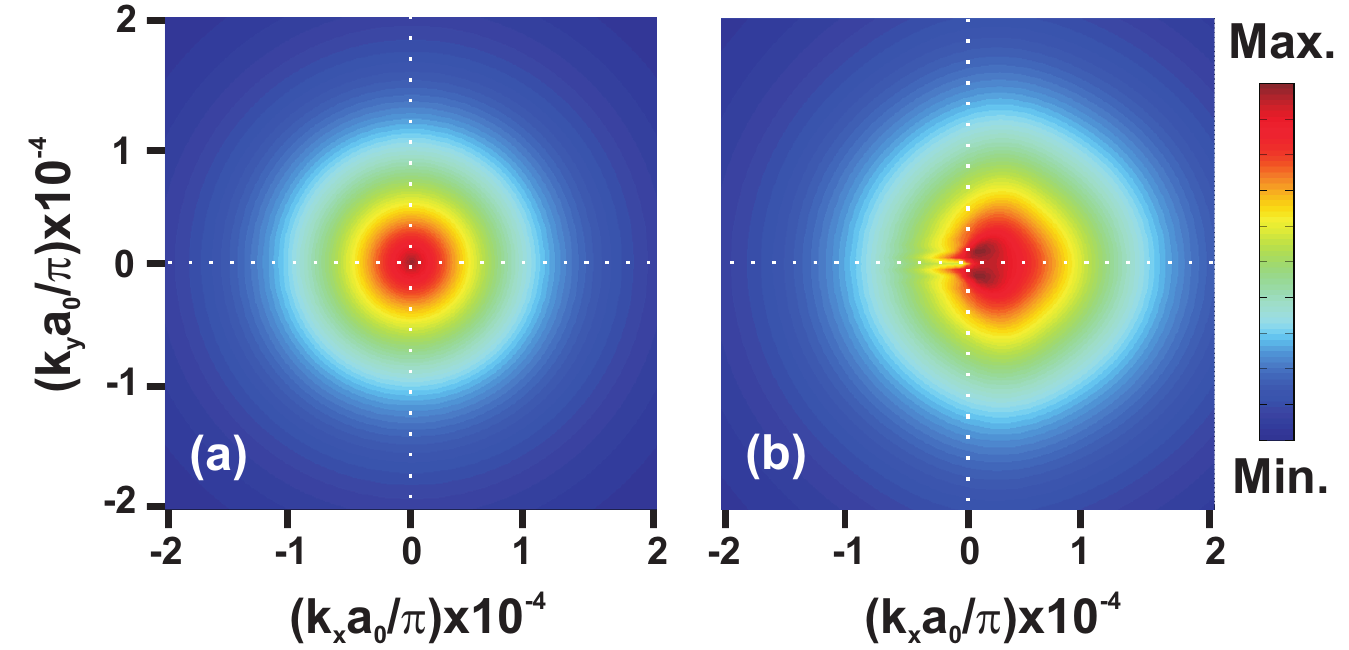}
\caption{(color online) Electron density distribution in $k$-space at (a)
the initial conditions and (b) at a time $t=$ 2.75 ps for a 100 V/cm
incident field. The white lines identify the position of the Dirac point.}
\end{figure}

To illustrate the combined effects of the interband and intraband
transitions on the carrier distributions, we plot the electron density in $k$%
-space in Fig. 3 before the pulse arrives (Fig. 3(a)) and at $t=2.75$ ps,
after almost half of the incident pulse has passed (Fig. 3(b)). At $t=2.75$
ps, the distribution has changed in two key respects: (i) the carriers have
been driven by the field so far to the right in $k$-space that most of them
now have a velocity component in the positive-$x$ direction, and (ii) the
carrier density is considerably increased just above and below the $k_{y}=0$
line just to the right of the Dirac point as expected from Eq.~(\ref{zeta}%
).\ The first effect is the source of clipping in the intraband current
density.\ The second effect results in the increase in carrier density,
which resulted in the increased intraband current in Fig. 2(c).\ However,
the most important effect of the strong redistribution of carriers in $k$%
-space is that it results in strong interplay between the interband and
intraband dynamics.\ Although it is tempting to interpret this interplay as
simply arising from the time-dependence of the Pauli-blocking of the states
near the Dirac point, as a physical picture based on a rate equation model
might suggest, such a picture does not lead to an accurate understanding of
the detailed current dynamics because all of the dynamics are occurring on a
sub-cycle timescale. Thus, the full simulation is
required.
\begin{figure}[t]
\includegraphics[width=8.5cm]{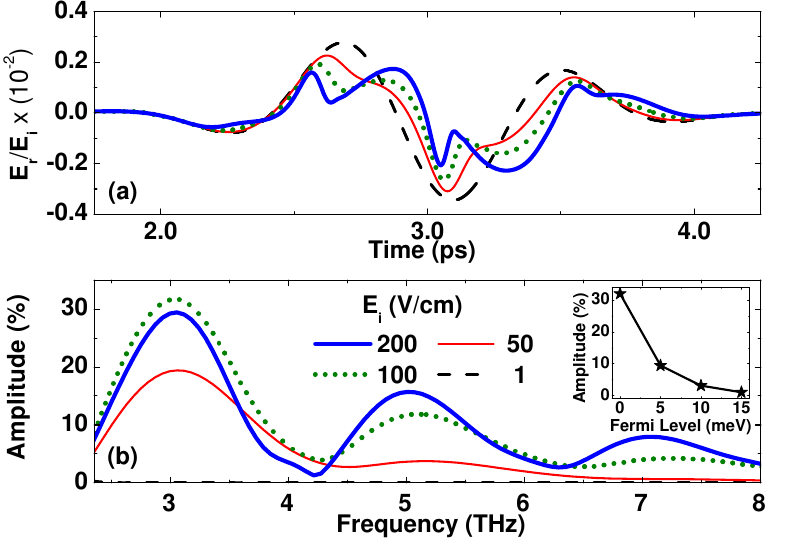}
\caption{(color online) (a) The reflected electric field normalized to the
corresponding amplitude of the incident electric field for four different
incident fields $E_{i}$, (b) amplitude spectra of the reflected signal
normalized to the peak at the fundamental frequency of 1 THz. The inset shows the dependence of the third harmonic amplitude as a function of the Fermi level for $E_{i}$= 200 V/cm.}
%Both plots are the result of the full calculation.}
\end{figure}

We now consider the transmitted and reflected THz fields. The transmitted field is given by $\mathbf{E}_{t}\left(t\right)=\mathbf{E}_{i}\left(t\right)-Z_{o}\mathbf{J}(t)/2$, where $Z_{o}$ is the free space impedance. Because the current density is relatively small for the field amplitudes considered, the transmitted field is dominated by the incident field. For this reason, we present the results for the reflected field, which is given by $\mathbf{E}_{r}\left(t\right)=-Z_{o}\mathbf{J}(t)/2$. Alternately, a differential transmission technique~\cite{Jiang2000} could be employed instead of the reflection configuration in order to measure the same response. The normalized time-dependent reflected fields for different field
amplitudes are shown for the full calculation in Fig. 4(a). The absolute peak of the un-normalized reflected%these two lines are new
field is 0.27 V/cm for the incident field of 100 V/cm.
As can be seen, the strong distortion of the interband current is
clearly exhibited in the reflected field. This distortion is an indication
of harmonics; the spectral responses normalized to the peak amplitude at the
fundamental frequency are presented in Fig. 4(b). While there is no harmonic
signal for the low field of 1 V/cm, the odd harmonics of the reflected field
emerge as the field amplitude increases. The third harmonic peaks at 32\% and 0.061\% of % new- Iadded the 0.06% in here.
the reflected and transmitted spectral peak at the fundamental, respectively for a 100 V/cm incident
field, and then decreases as the field is raised beyond that. 

We have also calculated the spectral response when
there is no interplay between the interband and intraband dynamics and find
(not shown) that the third harmonic amplitude is only 1.6\% and 0.6\%,
respectively when the interband and intraband dynamics are shut off.
Therefore, it is the interplay between the interband and intraband dynamics
that is crucial to the appearance of the strong harmonic generation in
undoped graphene. It is particularly noteworthy that we are obtaining a very
strong nonlinearity even though the carrier densities are not very high at
all; e.g., $n=2.24\times 10^{8}/cm^{2}$ for the 100 V/cm incident field. We note that in order to observe the strong nonlinearity, it is critical that the initial carrier density is low, otherwise Pauli-blocking effectively shuts off the interband transitions. This is demonstrated in the inset of Fig. 4(b), where we plot the third harmonic amplitude as a function of the Fermi energy. It is clearly evident that when the Fermi energy changes from 0 to 15 meV, the third harmonic amplitude is reduced from 32\% to 1\%. We have also performed simulations with zero Fermi energy but at room temperature and find that, again due to Pauli-blocking, the amplitude of the third harmonic generation is reduced by more than one order of magnitude. It is because all previous experiments on the nonlinear THz response of monolayer graphene have been performed on doped graphene ($|\mu_{F}|>$15 meV) or at temperatures well above 10 K, that the strong nonlinear effect that we predict has not been observed to date.

In order to investigate the origin of the decrease in the third harmonic amplitude beyond a certain field strength, we present in Fig. 5 the third, the fifth and the seventh harmonic amplitude versus the incident electric field for the results presented in Fig. 4(b). While the $5^{th}$ and the $7^{th}$ harmonic levels are negligible for fields up to 20 V/cm (as indicated by the arrow in Fig. 5), they start to increase gradually after that. At the same time a cubic fit (black curve) to the third harmonic departs from calculated third harmonic amplitude. This indicates that %this is new
the final third harmonic is not simply a third order process.
For example, at moderate fields it is due to the combination of a
$\chi^{3}$ process and a $\chi^{5}$ process that
contributes to the third harmonic, as shown diagrammatically in Figs.
6(a) and 6(b). This contribution of the fifth-order process to the third harmonic is
analogous to the Kerr effect for the linear response but at a higher order. That the reduction
in the third harmonic is initially due to a fifth order process is
evident in the fact that the fifth harmonic starts to appear at essentially
the same field as the third harmonic starts to deviate from the cubic fit. We have further
confirmed the origin of the decrease in the third harmonic by examining
the response when any self-fields are omitted and find essentially
the same trends in the response. As can also be seen from Fig. 5, higher order odd harmonics such as the $5^{th}$ and the $7^{th}$ also appear
as the field increases and reach levels that are 15.3\% and 7.7\% of the reflected fundamental, respectively.

\begin{figure}[t]
\includegraphics[width=7.cm]{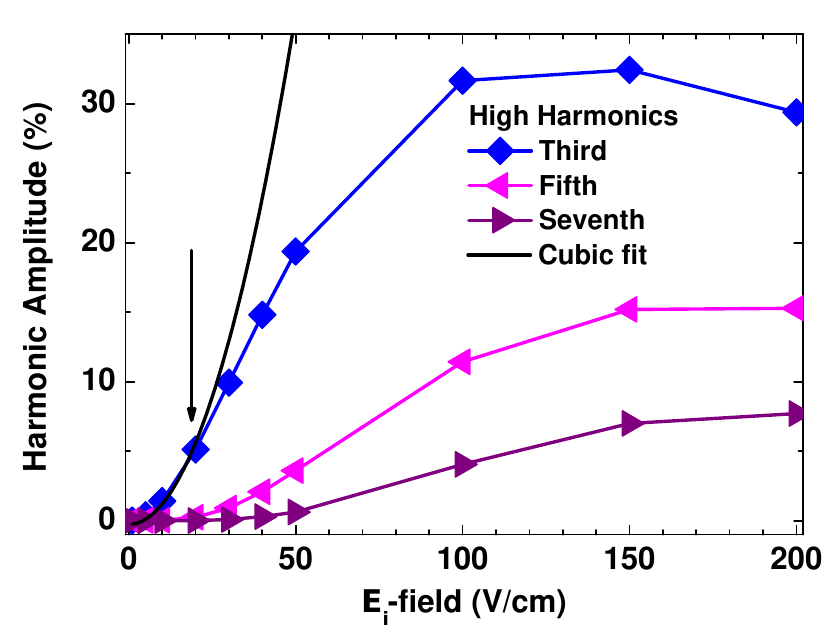}
\caption{(color online) The amplitude of the third, fifth and seventh harmonic normalized to the peak in the reflected field spectrum at the fundamental frequency at scattering time of 50 fs as a function of the incident field strength with the black curve shows a cubic fit.}
\end{figure}
\begin{figure}[t]
\includegraphics[width=5.5cm]{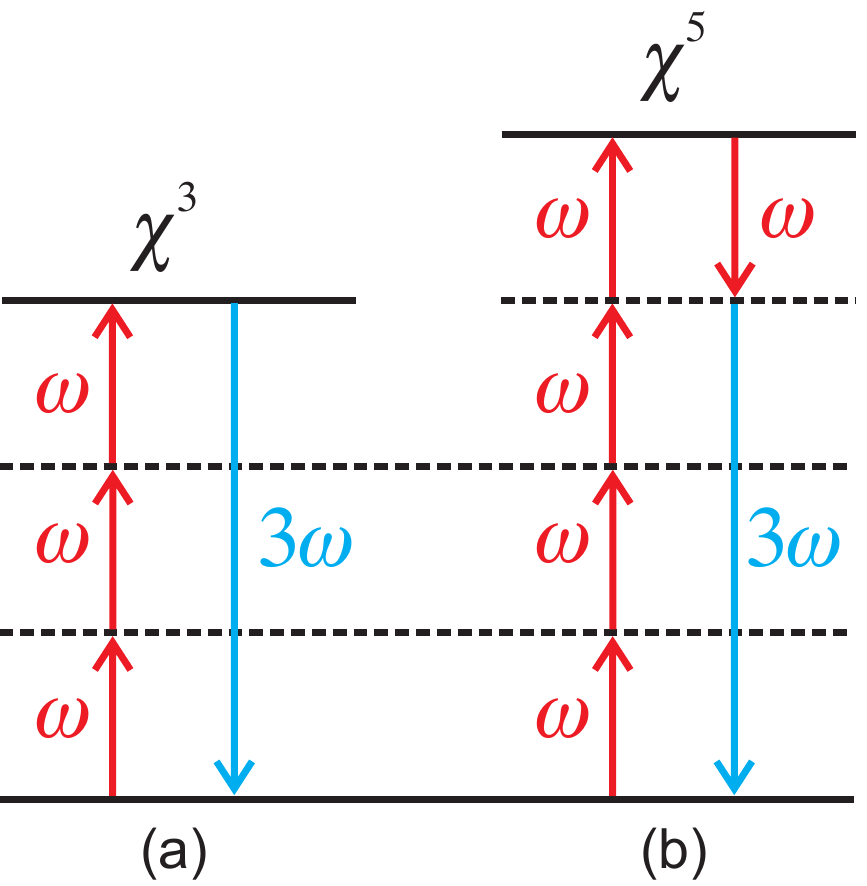}
\caption{(color online) Two possible mixing processes for $\chi^{3}$ susceptibility: (a)
direct $\chi^{3}$ process and (b) indirect $\chi^{3}$ process as
a result of $\chi^{5}$ process.}
\end{figure}

We now discuss how one could best experimentally observe the harmonic signals that we predict and what the noise requirements of the experimental set-up. First of all, in order to suppress any spectral components of the incident field that lie near the third harmonic, one should use a low-pass filter before the field reaches the graphene \cite{Paul2013}. After the signal has passed the sample, a high-pass filter at the detection side should ideally be used in order to remove the fundamental, so as to allow the measurement of the high harmonics only. Terahertz time-domain spectrometers (THz-TDS) have become an established method for sensitive measurements, and conventionally they are operated in a transmission-mode configuration. If the noise in the incident pulses in a transmission-mode experimental setup is found to be a limiting factor, a measurement technique called differential time-domain spectroscopy (DTDS) can be employed instead \cite{Jiang2000,Mickan2002,Liu2007,Withawat2007}. The reference signal will be transmission through air, and hence the final differential transmission measurements will be essentially identical to the reflection waveforms presented in Fig. 4(a). Using this technique, up to two orders of magnitude improvement in noise performance can be achieved \cite{Withawat2007}. The dynamic range of THz spectrometers is a very important factor to consider \cite{Mira2013}. It is defined as the ratio of the frequency dependent signal strength to the detected noise floor \cite{Jepsen2005,Piesiewicz2007}. One example of a low noise system has been achieved by Zhao et al. using a laser oscillator with 15 fs pulses and a 72 MHz repetition rate \cite{Zhao2002} and a semi-large photoconductive antenna at the emitter side. This setup achieved a dynamic range of 118 dB using a standard electro-optic detection scheme with a 1 mm thick (110) oriented ZnTe crystal \cite{Zhao2002}. Moreover, in this system, a field strength on the order of 95 V/cm was generated, which is sufficient to observe the predicted harmonics. Other researchers have employed a multi-scan technique that can be employed in order to increase the dynamic range. Recently, an ultra-high dynamic range of 90 dB was reported with 1000 scans that were taken in less than a minute \cite{Vieweg2014}. The third harmonic level for an incident field strength of 100 V/cm is predicted by our calculations to be 0.061\% in the transmission mode, i.e., 64.3 dB less than the fundamental. Hence, one should be able to measure the third harmonic with the experimental techniques and setups described above, as the field amplitudes are high enough and the achieved dynamic ranges of 90 dB and 118 dB are much larger than the required dynamic range of approximately 70 dB. 

We finally turn to the effect of scattering time on the generated third harmonic. As mentioned in the introduction, the effect of the scattering time on the response of graphene is particularly important in the THz regime. In Fig. 7, we now present the normalized third harmonic as a function of incident THz field amplitude for scattering times ranging from 10 fs to 100 fs. Because scattering reduces coherences in the carrier dynamics, it is expected that for a given field strength, larger scattering times will result in larger amplitudes of the third harmonic component of the transmitted and reflected fields. Our results indeed support this hypothesis as shown in Fig. 7.
\begin{figure}[t]
\includegraphics[width=7.cm]{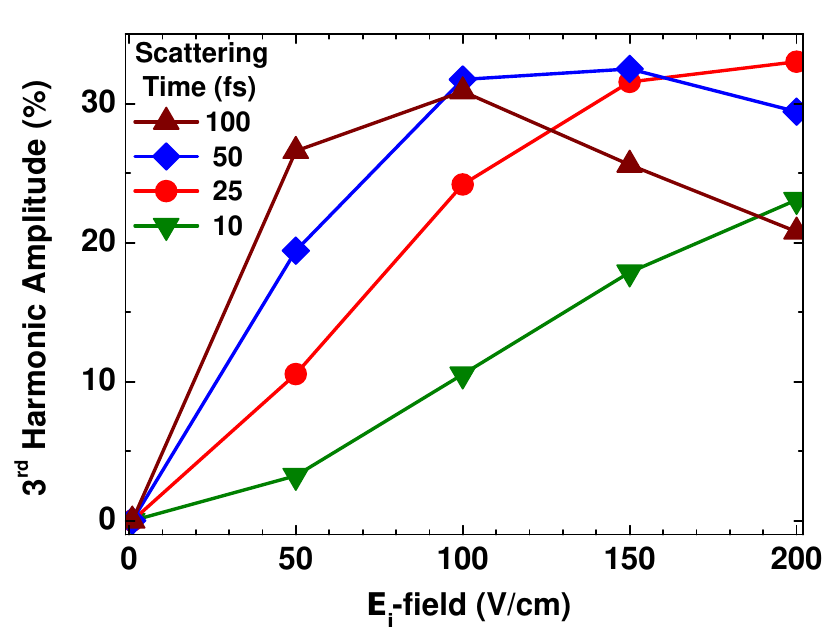}
\caption{(color online) Third harmonic amplitude of the reflected signal normalized to the peak in the reflected field spectrum at the fundamental frequency vs. the incident field strength with scattering time constants of 10, 25, 50 and 100 fs.}
\end{figure}
When $\tau$ is only 10 fs, the movement of the carriers in $k$-space is almost negligible as $\nu\times\tau=0.01$, which results in the carriers relaxing very quickly towards a thermal distribution around the Dirac point before they can move far from their equilibrium positions in $k$-space. This leads to a reduction in the interplay of the intraband and interband dynamics and thus a reduction in the nonlinearity. As the scattering time increases, the required electric field to induce the same displacement of the carriers becomes smaller. Thus, when the scattering time is increased by one order of magnitude to 100 fs, where $\nu\times\tau=0.1$, the carriers can be driven much further from the Dirac point. This results in a periodic reduction in Pauli-blocking of states near the Dirac point, that allows for a periodic increase in the interband response at these times, which in turn yields a strongly nonlinear response in the interband current density. We note that although an increase in the scattering time results in an increase in the third harmonic for fields up to about 100 V/cm (for the range of scattering times considered), due to higher-order effects, the maximum third harmonic level is essentially independent of the scattering time. However, the incident field amplitude at which the maximum occurs is strongly dependent on the scattering time. This means that our results are robust to reductions in the scattering time down to at least 10 fs.

\section{Summary}
In conclusion, we have theoretically investigated nonlinear high harmonic
generation in monolayer graphene. Our results demonstrate a very strong
interplay between the intraband and interband dynamics, leading to large odd
harmonics in the reflected field from suspended undoped graphene at low
temperature. This work lays out conditions under which future
experiments could achieve efficient high harmonic
generation in monolayer graphene at low density where electron-electron scattering is expected to have a minimal effect.

We thank the Natural Sciences and Engineering Research Council of Canada for
financial support.
\end{document}